\documentclass[aps,prb,reprint,amsmath,amssymb,superscriptaddress]{revtex4-2}
\usepackage[svgnames,psnames]{xcolor}
\usepackage{graphicx,hyperref,upgreek}
\usepackage[thicklines]{cancel}

\hypersetup{
colorlinks=true,
citecolor=NavyBlue,
linkcolor=DarkRed
}
\usepackage{bm,ulem}
\usepackage{color}

\begin{document}

\title{Cooperative Nernst Effect of Multilayer Systems: Parallel Circuit Model Study }
\author{Hiroyasu Matsuura }
\email{matsuura@hosi.phys.s.u-tokyo.ac.jp}
\affiliation{Department of Physics, University of Tokyo, 7-3-1 Hongo, Bunkyo, Tokyo 113-0033, Japan}

\author{Alexander Riss}
\author{Fabian Garmroudi}
\author{Michael Parzer}
\author{Ernst Bauer}
\affiliation{Institute of Solid State Physics, Technische Universit\"{a}t Wien, Vienna, Austria}

\date{\today}

\begin{abstract} 
Transverse thermoelectric power generation has emerged as a topic of immense interest in recent years owing to the orthogonal geometry which enables better scalability and fabrication of devices.
Here, we investigate the thickness dependence of longitudinal and transverse responses in film-substrate systems i.e., the Seebeck coefficient, Hall coefficient, Nernst coefficient and anomalous Nernst coefficient in a unified and general manner based on the circuit model, which describes the system as the parallel setup. 
By solving the parallel circuit model, we show that the transverse responses exhibit a significant peak, indicating the importance of a cooperative effect between the film and the substrate, arising from circulating currents that occur in these multilayer systems in the presence of a temperature gradient.
Finally, on the basis of realistic material parameters, we predict that the Nernst effect in bismuth thin films on doped silicon substrates is boosted to unprecedented values if the thickness ratio is tuned accordingly, motivating experimental validation. 
\end{abstract}

\maketitle

\section{Introduction}
The Seebeck and Nernst effect, which convert a temperature gradient into a longitudinal and transverse voltage, respectively, have attracted much attention in recent years, not only in terms of fundamental science but also from the perspective of effective utilization of thermal energy \cite{2020Nandihalli,2019Mizuguchi}.
Heavily doped semiconductors have been studied since the middle of the 20th century, while other classes of materials such as semimetallic systems including Heusler compounds \cite{2019Hinterleitner,2022Garmroudi,2023Bourgault,2023Fujimoto}
or Dirac and Weyl semimetals \cite{2018Skinner,2022Hosoi,2022Mizoguchi} as well as transition metal systems with strong interband scattering \cite{2023Garmroudi} have also been the focus of recent attention.
In particular, research on thermoelectric materials focusing on drag effects such as magnon drag \cite{2015Ang,2021Matsuura} and paramagnon drag effects \cite{2019Tsujii,2022Endo} for longitudinal thermoelectric response and the anomalous Nernst effect using exotic magnetic states for transverse thermoelectric effect have been in a specific focus of research \cite{2019Mizuguchi}.

With the development of microfabrication technology, thermoelectric effects in film-substrate systems have been also studied extensively \cite{2003Delatorre,2017Yordanov,2019Shimizu,2019Zhang,2019Byeon}.
For example, regarding the Seebeck effect, it was discovered that FeSe films on a SrTiO$_3$ substrate exhibit huge power factors \cite{2019Shimizu,2023matsubara}.
Furthermore, it has been reported that MoO$_{2+x}$ shows a much larger power factor on silicon (Si) than on quarz substrate \cite{2019Zhang}.
There have also been various studies on the transverse thermoelectric effect in film-substrate systems \cite{2000Young,2002Young,2003Young,2020Park,2017Chuang,2018Hu,2020Park,2022Yamazaki,2021Zhou,2021Yamamoto,2022Zhou,2023Zhou}.
The ordinary Nernst effect in such systems has been investigated as an example of transport properties in semiconducting films with a perspective of application \cite{2000Young,2002Young,2003Young}. On the other hand, a large anomalous Nernst effect has been discovered in ferromagnetic films, resulting in active research on the anomalous Nernst effect in film-substrate systems \cite{2017Chuang,2018Hu,2020Park,2022Yamazaki}. In full-Heusler Co$_2$MnGa films, a very large value of about -3$\upmu$V/K has been observed at a thickness of 40 nm \cite{2020Park}. It has also been reported that the anomalous Nernst coefficient increases with decreasing thickness in ferromagnetic Fe or Ni films on Si substrates \cite{2017Chuang}.

The above-indicated studies infer the importance of a substrate contribution to measurements in the film-substrate system. 
Presently, phenomenological theoretical methods for discussing the influence of the substrate on the Seebeck coefficient of film-substrate systems \cite{2015Quintana,2023Riss} and those for two-material composite \cite{1991Bergman,1999Bergman,2024Riss} have been discussed.

More recently, in bilayer systems of Fe-Ga films on n-type silicon substrate, it was found that the anomalous Nernst effect unexpectedly increases drastically as the thickness of substrate increases, reaching a peak of 15 $\upmu$V/K, being much larger than that of bulk Fe-Ga and much larger than the current record observed in Weyl semimetal Co$_2$MnGa \cite{2024Zhu}. 
The thickness dependence of the bilayer system has been also studied on the basis of a circuit model, and theoretical results are in good agreement with experiments \cite{2024Zhu}.  

Clearly, the film-substrate system has the potential to originate a huge thermoelectric effect by tuning the thickness.
Therefore, in this paper, we discuss the Hall coefficient, Seebeck effect and transverse thermoelectric effects such as normal and anomalous Nernst effects in film-substrate systems in a unified and general manner based on the parallel circuit model in matrix form. 
In conclusion, the Seebeck coefficients vary gradually with parameters characterizing the ratio of the thickness of film and substrate, while it is found that both the anomalous Nernst effect (studied in ref.\ \cite{2024Zhu}) as well as the ordinary Nernst effect exhibit a significant peak.
Finally, as an realistic example of unsuspected features, we study the Nernst effect in bismuth film on doped silicon substrate system.

The paper is organized as follows. In Sec. II, we introduce the linear response for the electrical and thermoelectric effects. In Sec.\ III, we derive the linear response in the film-substrate system based on the circuit model, which is used in Sec. IV to perform model calculations. In Sec.\ V, we illustrate the results with bismuth films on silicon substrate.

\section{Linear response for the electrical and thermoelectric effects}
In this section, we study the linear response for electrical and thermoelectric effects with/without magnetic field \cite{Ziman}.
When applied fields in the system A comprise an electric field (${\bm E}_{\rm A}$), a thermal gradient ($({\nabla} T)_{\rm A}$), and a magnetic field along the $z$-axis ($B_{z,{\rm A}}$), the linear response in the $x$,$y$ directions of system A is
\begin{eqnarray}
\begin{pmatrix}
j^x_{\rm e,A}  \\
j^y_{\rm e,A}  \\
j^x_{\rm Q,A}  \\
j^y_{\rm Q,A}  \\
\end{pmatrix}
= \bar{L}_{\rm A} 
\begin{pmatrix}
E_{x, {\rm A}}  \\
E_{y,{\rm A}}  \\
-(\nabla T)_{x,{\rm A}}/T  \\
-(\nabla T)_{y,{\rm A}}/T  \\
\end{pmatrix}  
,\label{eq_t1}
\end{eqnarray}
where ${\bm j}_{\rm e} =(j^x_{\rm e,A} , j^y_{\rm e,A} )$ and ${\bm j}_{\rm Q}=(j^x_{\rm Q,A} , j^y_{\rm Q,A} )$ are electrical and thermal current densities in system A, respectively.
$\bar{L}_{\rm A}$ is given by
\begin{eqnarray}
\bar{L}_{\rm A} = 
\begin{pmatrix}
\bar{L}_{11,{\rm A}} & \bar{L}_{12,{\rm A}} \\
\bar{L}_{21,{\rm A}} & \bar{L}_{22,{\rm A}}  \\
\end{pmatrix}
,\label{eq_t2}
\end{eqnarray}
where $\bar{L}_{ij,{\rm A}}$ is a $2 \times 2$ matrix of system A represented by
\begin{eqnarray}
\bar{L}_{ij,{\rm A}} = 
\begin{pmatrix}
L_{ij,{\rm A}}^{xx} & L_{ij,{\rm A}}^{xy} \\
L_{ij,{\rm A}}^{yx} & L_{ij,{\rm A}}^{yy}  \\
\end{pmatrix}
.\label{eq_t3}
\end{eqnarray}
In this paper $L_{ij,{\rm A}}^{kl}$ is called the linear response coefficient of system A, and $L_{ij,{\rm A}}^{xy} = -L_{ij,{\rm A}}^{yx}$.
In particular, $L_{11}^{xx (yy)}$ and $L_{11}^{xy}$ are the same as $\sigma^{xx (yy)}$ and $\sigma^{xy}$, which are the electrical conductivity and the Hall conductivity, respectively.    
As shown next, the Hall coefficient, Seebeck coefficient, Nernst coefficient and anomalous Nernst coefficient, which are transverse thermoelectric effects with/without a magnetic field, can be expressed in terms of the linear response coefficients.

\subsection{Hall coefficient}
The Hall effect measures the electric field that appears in the $y$ direction when an electrical current is applied in the $x$ direction in the presence of a perpendicular magnetic field $B_{z,{\rm A}}$ and in the absence of a thermal gradient.
The Hall coefficient of system A ($R_{\rm H,A}$) is defined as
\begin{eqnarray}
R_{\rm H,A} = \frac{E_{y,{\rm A}}}{j_{e,{\rm A}}^{x} B_{z,{\rm A}} }.
\end{eqnarray}
From Eqs.\ (\ref{eq_t1})-(\ref{eq_t3}),  we obtain
\begin{eqnarray}
\begin{pmatrix}
j^x_{\rm e,A}  \\
j^y_{\rm e,A}  \\
\end{pmatrix}
= 
\bar{L}_{11,{\rm A}} 
\begin{pmatrix}
E_{x,{\rm A}}  \\
E_{y,{\rm A}}  \\
\end{pmatrix}
. \label{hall_2}
\end{eqnarray}
By solving Eq.\ (\ref{hall_2}) in the condition of $j^y_{\rm e,A}=0$, the Hall coefficient is given as
\begin{eqnarray}
R_{\rm H,A} = \frac{L_{11,{\rm A}}^{xy}}{L_{11,{\rm A}}^{xx}L_{11,{\rm A}}^{yy} B_{z,{\rm A}}}. \label{Hall_3}
\end{eqnarray}

\subsection{Seebeck and Nernst coefficients}
The Seebeck and Nernst coefficients are obtained by measuring the electric field that appears in the $x$ and $y$ direction, respectively, when a thermal gradient is applied in the $x$ direction, and, for the Nernst coefficient, a magnetic field is present.
The Seebeck coefficient $S_{xx,\text{A}}$ and Nernst coefficient $\nu_{yx,\text{A}}$ are defined as
\begin{eqnarray}
S_{xx,{\rm A}} &=&\frac{E_{x,{\rm A}}}{(\nabla T)_{x,{\rm A}}},  \label{Seebeck_0} \\
\nu_{yx,{\rm A}} &=& \frac{E_{y,{\rm A}}}{(\nabla T)_{x,{\rm A}} B_{z,{\rm A}}}.
\end{eqnarray}

If ${\bm j}_e$ is zero, the linear response equation reads
\begin{eqnarray}
 0 =  \bar{L}_{11,{\rm A}} 
 \begin{pmatrix}
E_{x,{\rm A}}  \\
E_{y,{\rm A}}  \\
\end{pmatrix}
 + 
  \bar{L}_{12,{\rm A}} 
 \begin{pmatrix}
-(\nabla T)_{x,{\rm A}}/T  \\
-(\nabla T)_{y,{\rm A}}/T  \\
\end{pmatrix}
.
\end{eqnarray}
Since the thermal gradient in the $y$ direction can be neglected,
\begin{eqnarray}
 \begin{pmatrix}
E_{x,{\rm A}}  \\
E_{y,{\rm A}}  \\
\end{pmatrix}
=- \bar{L}_{11,{\rm A}}^{-1} \bar{L}_{12,{\rm A}}
 \begin{pmatrix}
-(\nabla T)_{x,{\rm A}}/T  \\
0  \\
\end{pmatrix}
,  \label{S_N_1}
\end{eqnarray}
where $\bar{L}_{11,{\rm A}}^{-1}$is the inverse matrix of $\bar{L}_{11,{\rm A}}$.

Ignoring the term containing $L_{11,{\rm A}}^{xy}L_{11,{\rm A}}^{yx}$, a second-order effect of the magnetic field, yields the Seebeck and Nernst coefficients as
\begin{eqnarray}
S_{xx,{\rm A}} &=& \frac{L_{12,{\rm A}}^{xx}}{TL_{11,{\rm A}}^{xx}},  \label{Seebeck_1} \\
\nu_{yx,{\rm A}} &=& \frac{ L_{11,{\rm A}}^{xy}L_{12,{\rm A}}^{xx} - L_{11,{\rm A}}^{xx}L_{12,{\rm A}}^{xy} }{TL_{11,{\rm A}}^{xx}L_{11,{\rm A}}^{yy}B_{z,{\rm A}} } \nonumber \\
&=& (S_{xx,{\rm A}} \theta_{{\rm H},{\rm A}} - S_{yy,A} \alpha_{{\rm N}, {\rm A}})/B_{z,{\rm A}},
 \label{Nernst_1} 
\end{eqnarray} 
where $\nu_{yx,{\rm A}} = -\nu_{xy,{\rm A}}$. $\theta_{{\rm H},{\rm A}}$ and $\alpha_{{\rm N},{\rm A}}$ are the Hall and Peltier angles \cite{2018Hu} defined as (See Appendix A)
\begin{eqnarray}
\tan{\theta_{{\rm H},{\rm A}}} \simeq \theta_{{\rm H},{\rm A}} &=&
\frac{L_{11,{\rm A}}^{xy}}{L_{11,{\rm A}}^{yy}}, \label{theta_1}\\
\tan{\alpha_{{\rm N}, {\rm A}}}  \simeq \alpha_{{\rm N}, {\rm A}} &=& \frac{L_{12,{\rm A}}^{xy}}{L_{12,{\rm A}}^{yy}}.  \label{alpha_1}
\end{eqnarray}
In the Nernst effect,  $L_{11, A}^{xy}$ is proportional to the magnetic field, i.e., $L_{11, A}^{xy} \propto B_{z,{\rm A}}$. 
Thus, using Eq.\ (\ref{Hall_3}), the Hall angle is given by
\begin{eqnarray}
\theta_{{\rm H},{\rm A}} = R_{\rm H,A}L_{11,{\rm A}}^{xx}B_{z,{\rm A}}.
\end{eqnarray} 

\subsection{Thermoelectric transports with anomalous components}
In the above discussion, we studied the Hall, Seebeck and Nernst coefficients.
On the other hand, $L^{xy}_{ij}$ can be finite in ferromagnetic materials even without a magnetic field, yielding to $\textit{anomalous}$ effects.

The anomalous Nernst coefficient is defined as 
\begin{eqnarray}
N_{yx,A} = \frac{E_{y,A}}{(\nabla T)_{x,A}} \label{AN}.
\end{eqnarray}
Using Eqs.\ (\ref{Seebeck_0}), (\ref{S_N_1}) and (\ref{AN}), the Seebeck coefficient with anomalous components and the anomalous Nernst coefficient are
\begin{eqnarray}
S_{xx,A}^{AN} &=& \frac{ L_{11,A}^{yy}L_{12,A}^{xx} + L_{11,A}^{xy}L_{12,A}^{xy} }{T(L_{11,A}^{xx}L_{11,A}^{yy} + (L_{11,A}^{xy})^2)}\nonumber \\ &=& \frac{S_{xx,A} + \theta_{\rm H, A} \alpha_{\rm N, A} r_{12} S_{yy,A}}{1 + \theta_{\rm H,A}^2 r_{11} } ,   \label{A_Seebeck_1} \\
N_{yx,A} &=& \frac{ L_{11,A}^{xy}L_{12,A}^{xx} - L_{11,A}^{xx}L_{12,A}^{xy} }{T(L_{11,A}^{xx}L_{11,A}^{yy} + (L_{11,A}^{xy})^2)} \nonumber \\ &=&  \frac{S_{xx,A}\theta_{\rm H,A} - S_{yy,A}\alpha_{\rm N, A}}{1 + \theta_{\rm H,A}^2 r_{11} } , \label{A_Nernst_1}
\end{eqnarray}
where $r_{ij} = L_{ij}^{yy}/L_{ij}^{xx}$ and $N_{yx,A} = - N_{xy,A}$.

\section{Linear response of a film-substrate system based on the circuit model}
\begin{widetext}
\begin{figure*}[t]
\begin{center}
\rotatebox{0}{\includegraphics[angle=0,width=1\linewidth]{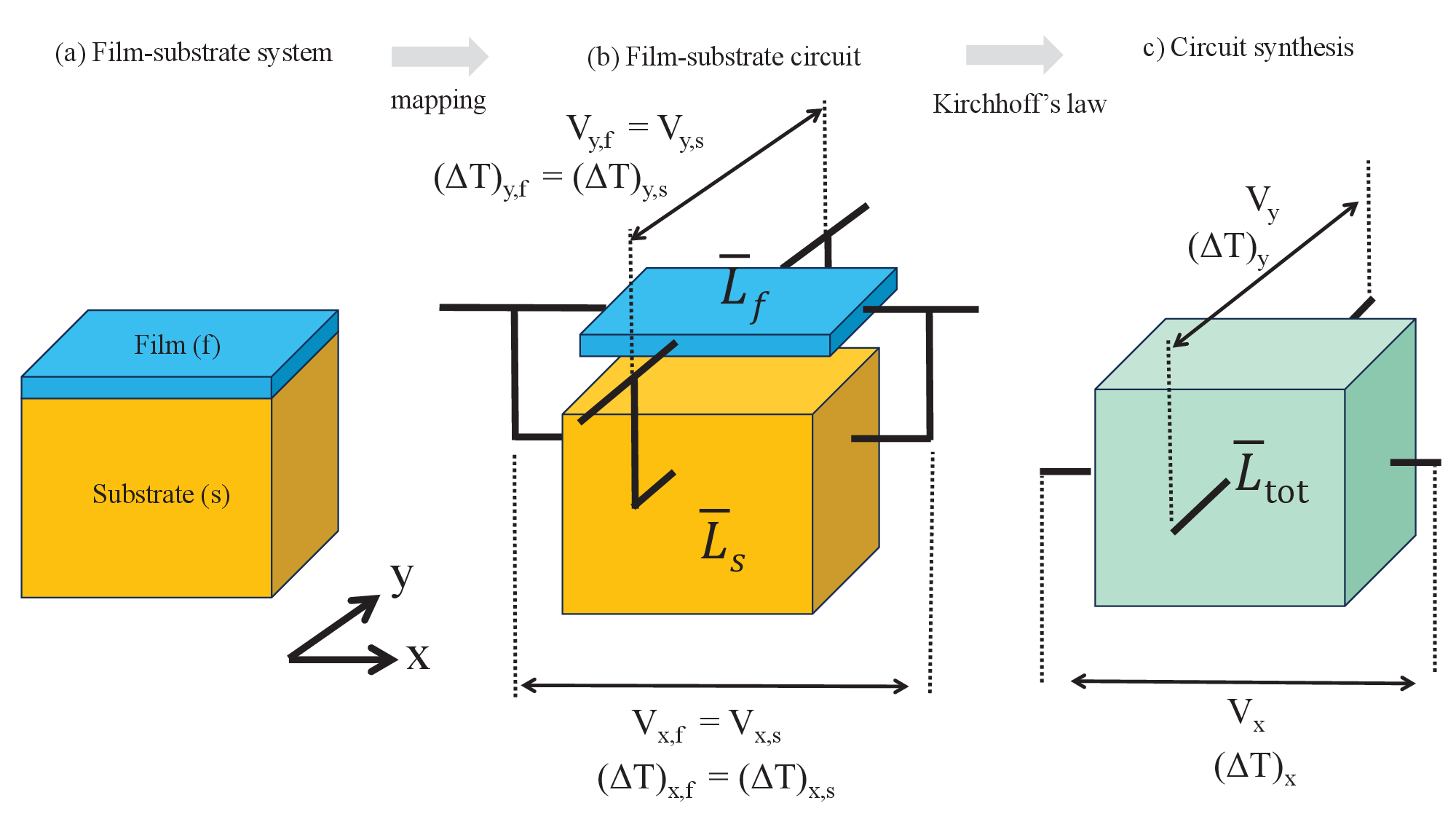}}
\caption{Schematic image of  a) film-substrate system, b) parallel circuit consisting of $\bar{L}_f$ and $\bar{L}_s$, and c) the circuit organized by Kirchhoff's law.  $V_{x,i} = E_{x,i} a$, and $(\Delta T)_{y,i} = (\nabla T)_{x,i} a$ for $i =s, f$, with $s$ and $f$ denoting the substrate and film, respectively.}
\label{Fig1}
\end{center}
\end{figure*}
\end{widetext}

In this section, we derive a linear response coefficient in the film-substrate system based on circuit model which treats the film and substrate as a parallel setup.
Fig.\ \ref{Fig1}(a) shows a sketch of a film-substrate systems.
As discussed in \cite{2023Riss}, this system can be mapped to the parallel circuit as shown in Fig.\ \ref{Fig1}(b). 
Here, the linear response coefficients of the film (system A $= f$) and substrate (A $= s$) are defined as $\bar{L}_f$ and $\bar{L}_s$, respectively.
We assume that $B_{z,f} = B_{z,s} \equiv B_{z}$.

In this paper, the volume of film and substrate are $v_f = a_x \times a_y \times d_f$ and $v_s = a_x \times a_y \times d_s$, respectively, where $a_{x (y)} $ is the length in $x$ $(y)$ direction, and $d_f$ and $d_s$ are thicknesses of film and substrate, respectively. 
From Kirchhoff's law in parallel circuits (see Fig.\ \ref{Fig1}(c)), the electrical potentials of the film and substrate ($V_{x,f}=E_{x,f} a_{x}$ and $V_{x,s}=E_{x,s} a_x$) satisfy the following equations:
\begin{eqnarray}
&V_{x,f} = V_{x,s} = V_{x}, \label{eq_ks_1}\\
&V_{y,f} = V_{y,s} = V_{y}, \label{eq_ks_2}
\end{eqnarray}
where $V_{x}$ and $V_{y}$ are electrical potentials in the $x$ and $y$ directions in the system of the combined circuit, respectively.
Thus, the electric fields in the $x$- and $y$- directions in the combined circuit are defined as $E_{x} = V_{x}/a_x$ and $E_{y} = V_{y}/a_y$, respectively.

The temperature gradient is considered similarly as for in the electrical circuit.
When the temperature differences in the $x$ ($y$) direction at film and substrate are $(\Delta T)_{x(y), f} =(\nabla T)_{x(y),f} a_{x (y)}$ and $(\Delta T)_{x(y), s} = (\nabla T)_{x(y),s} a_{x (y)}$, we obtain
\begin{eqnarray}
&(\Delta T)_{x,f}= (\Delta T)_{x,s} = (\Delta T)_x,  \label{eq_ks_3}\\
&(\Delta T)_{y,f} = (\Delta T)_{y,s} = (\Delta T)_y,  \label{eq_ks_4}
\end{eqnarray}
where $(\Delta T)_x$ and $(\Delta T)_y$ are the temperature differences in the combined circuit.

In the parallel system formed by the film and the substrate, the total electrical ($\beta =e$) and thermal ($\beta =Q$) currents in $x$ (y) direction are given as
\begin{eqnarray}
j_{\beta,{\rm tot}}^{x (y)}a_{y (x)}(d_s + d_f) = j_{\beta,f}^{x (y)}a_{y (x)}d_f + j_{\beta,s}^{x (y)}a_{y (x)}d_s. \label{eq_ks_5}
\end{eqnarray}

From Eqs.\ (\ref{eq_t1}), (\ref{eq_ks_1})-(\ref{eq_ks_4}), and (\ref{eq_ks_5}), the total linear response coefficient of film-substrate system is given as
\begin{eqnarray}
\bar{L}_{\rm tot} \equiv \frac{d_f}{d_f + d_s}\bar{L}_{f} + \frac{d_s}{d_f + d_s}\bar{L}_{s}. \label{eq_sf_1}
\end{eqnarray}

For simplicity, we assume below that the response is isotropic in the $x$- and $y$-directions.
Applying Eq.\ (\ref{eq_sf_1}) to Eqs.\ (\ref{Seebeck_1}), (\ref{theta_1}) and (\ref{alpha_1}), the Seebeck coefficient ($S_{\rm tot}$), the Hall angle ($\theta_{\rm H,\rm tot }$), and the Peltier angle ($\alpha_{\rm N,\rm tot }$) are
\begin{eqnarray}
S_{\rm tot} &=& \frac{S_f + \epsilon_\sigma S_s}{1 + \epsilon_\sigma}, \label{Seebeck_tot} \\
\theta_{\rm H,\rm tot } & = & \frac{\theta_{\rm H, f } + \epsilon_\sigma \theta_{\rm H, s }}{1 + \epsilon_\sigma}, \label{theta_tot} \\
\alpha_{\rm N,\rm tot } & = & \frac{ S_f \alpha_{\rm N, f } +  \epsilon_\sigma S_s \alpha_{\rm N, s }  }{S_f + \epsilon_\sigma S_s  }, \label{alpha_tot} 
\end{eqnarray}
and 
\begin{eqnarray}
\epsilon_{\sigma} = \frac{L_{11,s}^{xx}d_s}{L_{11,f}^{xx}d_f} = \frac{L_{11,s}^{yy}d_s}{L_{11,f}^{yy}d_f} = \frac{R_f}{R_s},  \label{Riss_define}
\end{eqnarray}
where $R_A$ is the resistance of film ($A=f$) or substrate ($A=s$) \cite{2023Riss}. 
The expression of the Seebeck coefficient in Eq.\ (\ref{Seebeck_tot}) is the same as in a previous study \cite{2023Riss}.
The Hall angle (Eq.\ (\ref{theta_tot})) and the Peltier angle (Eq.\ (\ref{alpha_tot})) are functions of the same form as the Seebeck coefficient (Eq.\ (\ref{Seebeck_tot})), when $\epsilon_\sigma S_s/S_f$ is replaced by $\tilde{\epsilon}_{\sigma}$ in Eq.\ (\ref{alpha_tot}).

Next, applying Eq.\ (\ref{eq_sf_1}) to Eq.\ (\ref{Hall_3}), the Hall coefficient ($R_{\rm H, \rm tot}$) in the film-substrate systems is
\begin{eqnarray}
R_{\rm H, \rm tot} 
=\frac{R_{{\rm H}, f}+ \epsilon_\sigma R_{{\rm H}, s} }{1 + \epsilon_\sigma}  +\frac{\epsilon_{\sigma}}{(1+ \epsilon_{\sigma})^2} R_{{\rm H}, s} \delta R_{{\rm H}}, \label{Hall_tot}  
\end{eqnarray}
where 
\begin{eqnarray}
\delta R_{{\rm H}} &=& (\rho_{f} -\rho_{s} ) ( \theta_{\rm H, s} - \theta_{\rm H, f}  )/R_{{\rm H}, s}B.
\end{eqnarray}
Here, $\rho_{f}$ and $\rho_{s}$ are the electrical resistivities of film and substrate, respectively.

Finally, applying Eq.\ (\ref{eq_sf_1}) to Eqs.\ (\ref{Nernst_1}), (\ref{A_Seebeck_1}) and (\ref{A_Nernst_1}), the Nernst coefficient ($\nu_{\rm tot}$), the Seebeck coefficient with the anomalous components ($S_{\rm tot}^{AN}$), and the anomalous Nernst coefficient ($N_{\rm tot}$) are
\begin{widetext}
\begin{eqnarray}
\nu_{\rm tot} &=& \frac{\nu_{\rm f}+ \epsilon_\sigma \nu_{\rm s} }{1 + \epsilon_\sigma}  +\frac{\epsilon_{\sigma}}{(1+ \epsilon_{\sigma})^2} \nu_{\rm s} \delta \nu, \label{Nernst_tot}  \\
S_{\rm tot}^{AN} &=& \frac{1}{1 + \theta_{\rm H, tot}^2 } \bigr[ \frac{(1 + \theta_{\rm H, f}^2) S_{\rm f}^{AN}+ \epsilon_\sigma (1 + \theta_{\rm H, s}^2) S_{\rm s}^{AN} }{1 + \epsilon_\sigma}  + \frac{\epsilon_{\sigma}}{(1+ \epsilon_{\sigma})^2} S_{\rm s}^{AN} \delta S^{AN} \bigr] \nonumber \\
&\simeq& \frac{S_{\rm f}^{AN}+ \epsilon_\sigma S_{\rm s}^{AN} }{1 + \epsilon_\sigma} +\frac{\epsilon_{\sigma}}{(1+ \epsilon_{\sigma})^2} S_{\rm s}^{AN} \delta S^{AN} ,  \label{S_AN_tot} \\
N_{\rm tot} &=& \frac{1}{1 + \theta_{\rm H, tot}^2 } \bigr[ \frac{(1 + \theta_{\rm H, f}^2)  N_{\rm f}+ \epsilon_\sigma (1 + \theta_{\rm H, s}^2)  N_{\rm s} }{1 + \epsilon_\sigma} +\frac{\epsilon_{\sigma}}{(1+ \epsilon_{\sigma})^2} N_{\rm s} \delta N \bigr] \nonumber \\
&\simeq& \frac{N_{\rm f}+ \epsilon_\sigma N_{\rm s} }{1 + \epsilon_\sigma} +\frac{\epsilon_{\sigma}}{(1+ \epsilon_{\sigma})^2} N_{\rm s} \delta N, \label{N_tot} 
\end{eqnarray}
\end{widetext}
where 
\begin{eqnarray}
\delta \nu &=& 
(S_f -S_s)(\theta_{\rm H, s} - \theta_{\rm H, f}  )/\nu_{\rm s}B_z
, \label{del_nu} \\ 
\delta S^{AN} &=& (S_f\alpha_{N,f} -S_s\alpha_{N,s})(\theta_{\rm H, s} - \theta_{\rm H, f}  )/S_{\rm s}^{AN}, \\ 
\delta N &=& (S_f -S_s )(\theta_{\rm H, s} - \theta_{\rm H, f}  )/N_{\rm s}. 
\end{eqnarray}
In Eqs.\ (\ref{S_AN_tot}) and (\ref{N_tot}), we used $1 + \theta_{H, i}^2 \simeq 1$ for $i = {\rm tot}$, $s$, and $f$.
Therefore, Eqs.\ (\ref{Hall_tot}), and (\ref{Nernst_tot}) - (\ref{N_tot}) show the same $\epsilon_{\sigma}$ dependence.
The anomalous Nernst coefficient of Eq.\ (\ref{N_tot}) is the same as in ref. \cite{2024Zhu}, when we neglect the Hall angle of substrate.
 
In the Appendix B, we derive a $\epsilon_{\sigma}$ dependence of Hall resistivity.
Although the linear responses of a single film-substrate system is explained in this section, it is easy to extend to the n-layer systems. In appendix C, the respective formulation is described.
   

\section{Model Calculations }
In this section, we further elaborate the differences between two types: type-I (Eqs.\ (\ref{Seebeck_tot})-(\ref{alpha_tot})), and type-II  (Eq.\ (\ref{Hall_tot})) and (Eqs.\ (\ref{Nernst_tot})-(\ref{N_tot})).
Particularly, the $\epsilon_\sigma$ dependencies, i.e. the weighted contribution characterizing the ratio of the film to substrate resistances, are studied for the Seebeck coefficient (type-I) and the Nernst coefficient (type-II).

\subsection*{Type-I}
\begin{figure}[h]
\begin{center}
\rotatebox{0}{\includegraphics[angle=0,width=1\linewidth]{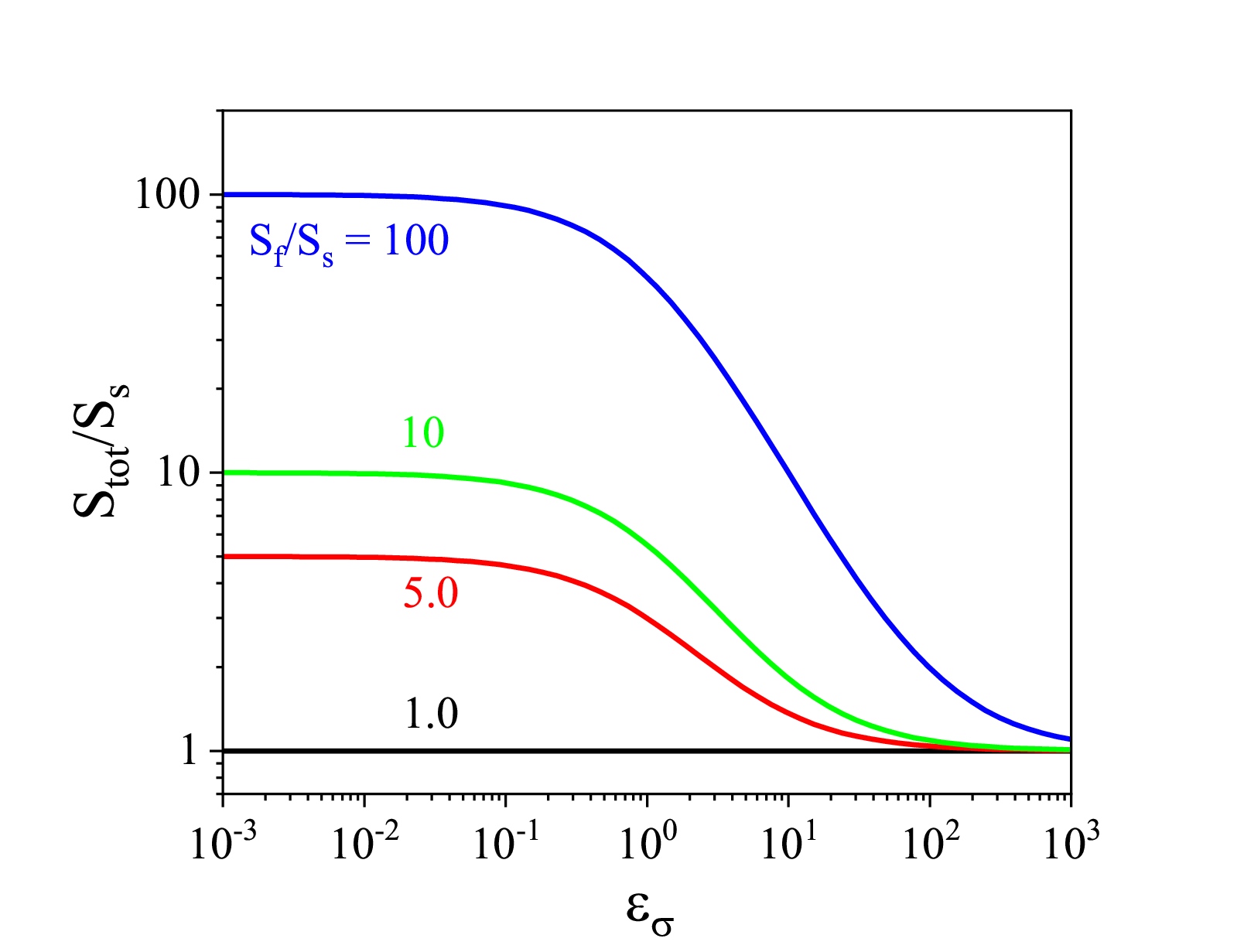}}
\caption{$\epsilon_{\sigma}$ dependence of the Seebeck coefficient (Type-I).  The Hall angle and the Peltier angle are the same forms of function as the Seebeck coefficient.}
\label{Fig_MC_Seebeck}
\end{center}
\end{figure}

Figure \ref{Fig_MC_Seebeck} shows the $\epsilon_{\sigma}$ dependence of the Seebeck coefficient.
The Seebeck coefficient gradually changes and there is no extremum.
The result is consistent with a previous study \cite{2023Riss}. 

The origin of this dependence is understood based on the circulating current shown in Fig. \ref{Fig_sche}(a) \cite{2023Riss}.
If a temperature difference is applied in the $x$ direction ($\Delta T$), a voltage is caused by both the Seebeck effect in the film and the substrate, respectively.
The circulating current $I_{circ}^x$ is flowing around the film and the substrate.
Employing Ohm's law, the circulating current reads
\begin{eqnarray}
I_{circ}^{x} = \frac{(S_f - S_s)\Delta T}{R_f + R_s}. \label{circ1}
\end{eqnarray}
Thus, the measured voltage in the $x$ direction is given by
\begin{eqnarray}
V_x &=& S_{f}\Delta T + I_{circ}^xR_f, 
\end{eqnarray}
and the measured Seebeck coefficient becomes
\begin{eqnarray}
S_{tot} = \frac{V_x}{\Delta T} = \frac{S_sR_s + S_fR_f}{R_s + R_f}.
\end{eqnarray}
This equation is consistent with Eq. (\ref{Seebeck_tot}).
The circulating current flowing around the film and substrate is thus essential to understand the $\epsilon_{\sigma}$ dependence of Eq.\ (\ref{Seebeck_tot}).

\subsection*{Type-II}
\begin{figure}[h]
\begin{center}
\rotatebox{0}{\includegraphics[angle=0,width=1\linewidth]{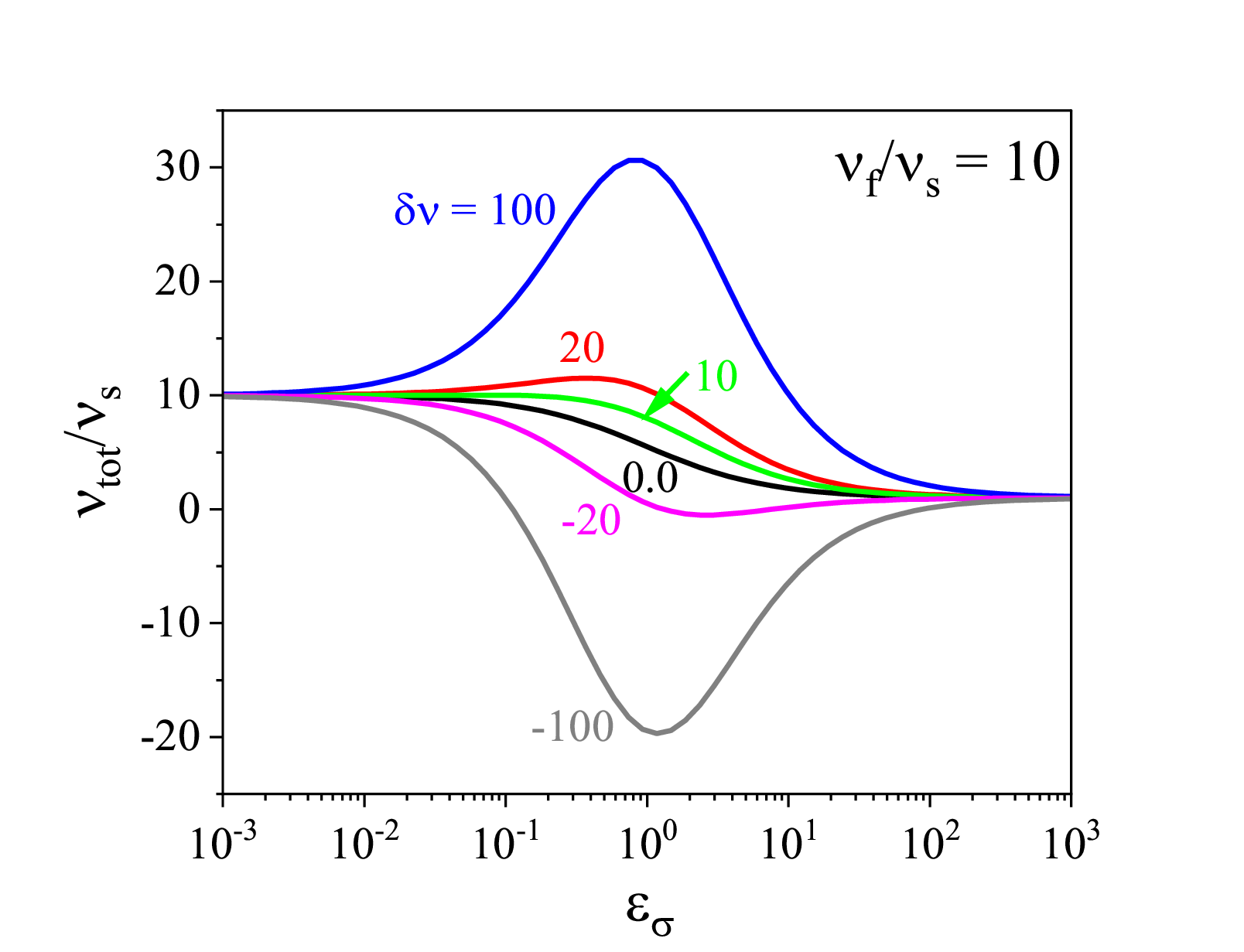}}
\caption{$\epsilon_{\sigma}$ dependence of the Nernst coefficient $\nu_{\rm tot}$ (Type-II).
We fix $\nu_{\rm tot}/\nu_{\rm s} = 10$. The Hall coefficient ($R_{\rm H, \rm tot}$), the Seebeck coefficient with the anomalous components ($S_{\rm tot}^{AN}$), and the anomalous Nernst coefficient ($N_{\rm tot}$) also show the same $\epsilon_{\sigma}$ dependence with the Nernst coefficient.}
\label{Fig_MC1_Nernst}
\end{center}
\end{figure}

Here, we study the type-II equations, such as the Nernst coefficient of Eq.\ (\ref{Nernst_tot}). 
If $\delta \nu = 0.0$ (see Eq.\ (\ref{del_nu})), $\nu_{tot}$ behaves the same as the Seebeck coefficient in Fig.\ \ref{Fig_MC_Seebeck}. 
On the other hand, when $\delta \nu$ is finite, an additional contribution appears.
As shown in Fig.\ \ref{Fig_MC1_Nernst}, it is found that the Nernst coefficient exhibits distinct maxima/minima for $\epsilon_{\sigma} \simeq 1.0 $ as $|\delta \nu|$ increases.

\begin{widetext}
\begin{figure*}[t]
\begin{center}
\rotatebox{0}{\includegraphics[angle=0,width=0.9\linewidth]{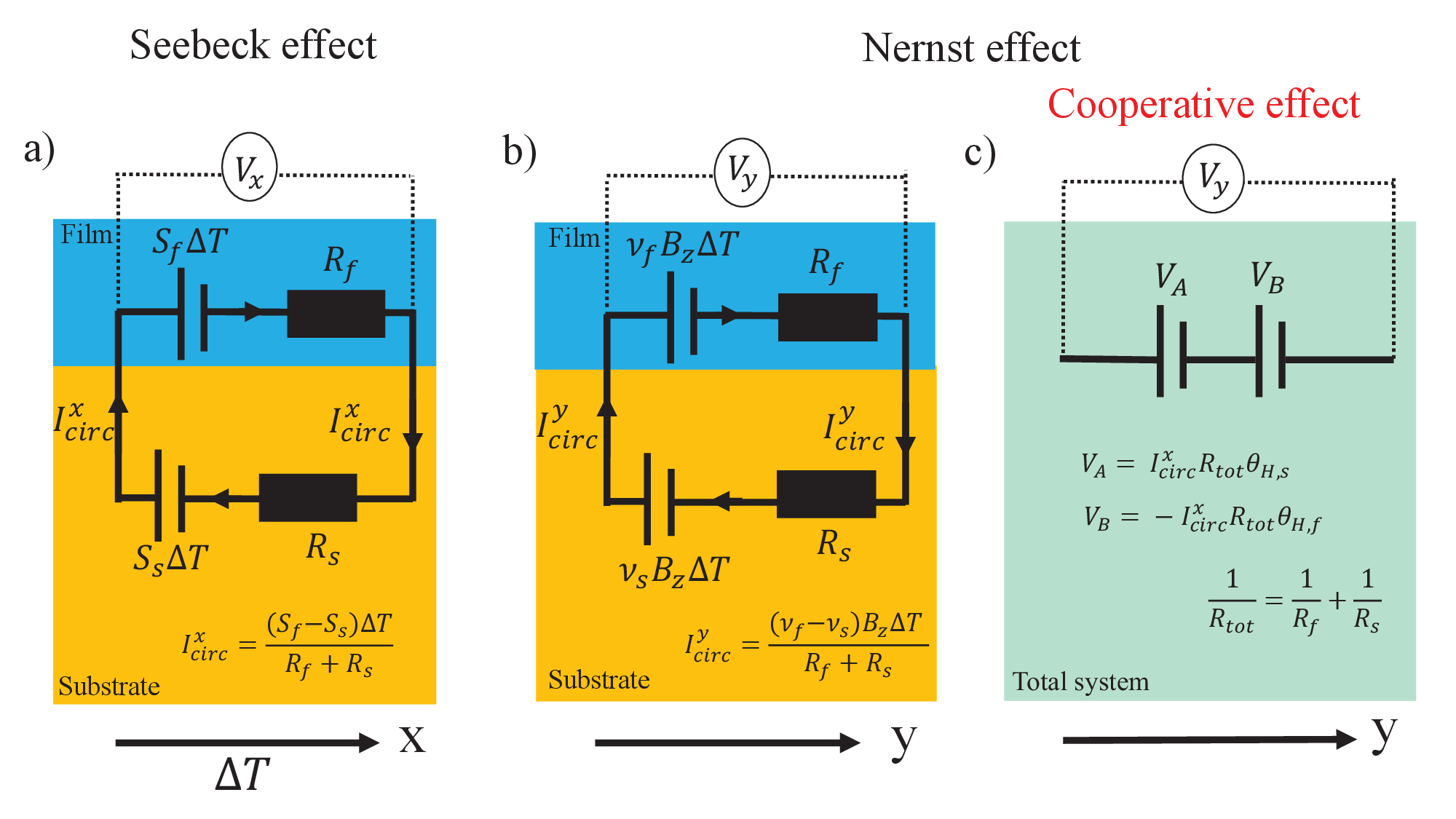}}
\caption{Schematic picture of the film-substrate system for describing the Seebeck effect (a) and Nernst effect (b)-(c), where $\Delta T$ is a temperature gradient in the x-direction. 
$I_{circ}^{x}$ is the circulating current flowing around the film and the substrate along x direction, and it is proportional to the difference of the Seebeck coefficients in the film and substrate. The total voltage is obtained by $V_x =I_{circ}^{x}(R_f + R_s)$, where $R_{f}$ and $R_s$ are resistances in the film and the substrate, respectively.  In the same way,  the circulating current along y direction ($I_{circ}^{y}$), which is proportional to the difference of the nernst coefficients in the film and the substrate, is flowing around the film and the substrate. As shown in (b), the nernst voltage is $V_y =I_{circ}^{y}(R_f + R_s)$.  {\rm The Cooperative effect} is the cross term of $I_{circ}^{x}$ and Hall angles in the film and the substrate ($\theta_{\rm H, s}$ and $\theta_{\rm H, f}$). The sign different between $V_A$ and $V_B$ shown in (c) is due to the direction of $I_{circ}^{x}$ in the film and the substrate. The voltage due to the cooperative effect is $V_y =V_A + V_B = I_{circ}^{x}R_{tot}(\theta_{\rm H, s} -\theta_{\rm H, f})$ where $R_{tot}$ is a resistance of the total system as defined by $1/R_{tot} = 1/R_f + 1/R_s$.
}
\label{Fig_sche}
\end{center}
\end{figure*}
\end{widetext}

To understand the origin of the $\epsilon_{\sigma}$ dependence, Eq.\ (\ref{Nernst_tot}) is divided into two terms as
\begin{eqnarray}
\nu_{\rm tot} &=& \nu_1 + \nu_2, \label{nu_divide}
 \end{eqnarray}
 where
\begin{eqnarray}
 \nu_1 &=& \frac{\nu_{\rm f}+ \epsilon_\sigma \nu_{\rm s} }{1 + \epsilon_\sigma}, \label{Nernst_tot1}\\
 \nu_2 &=& \frac{\epsilon_{\sigma}}{(1+ \epsilon_{\sigma})^2} (S_f -S_s)(\theta_{\rm H, s} - \theta_{\rm H, f}  )/B_z.  \label{Nernst_tot2}
\end{eqnarray}
If a temperature difference is applied in the $x$ direction and the magnetic field perpendicular to film-substrate systems, we conclude, in analogy to the Seebeck case (Fig.\ \ref{Fig_sche}(a)), for $I_{circ}^y$ (Fig.\ \ref{Fig_sche}(b))
\begin{eqnarray}
I^{y}_{circ} = \frac{(\nu_f -\nu_s) B_z \Delta T}{R_f + R_s};
\end{eqnarray}
the measured Nernst coefficient is then
\begin{eqnarray}
\frac{V_y}{B_z \Delta T } = \frac{\nu_f R_f  + \nu_s R_s}{R_f + R_s}, 
\end{eqnarray}
which is the same as $\nu_1$ (Eq.\ (\ref{Nernst_tot1})). 
Thus, the origin of $\nu_1$ is caused by the circulating current in the y-direction, flowing the film and the substrate.

Next, let us discuss the origin of $\nu_2$.
Using the circulating current of Eq.\ (\ref{circ1}) ($I_{circ}^{x}$) and the resistance of the total system $R_{tot}$ defined by $1/R_{tot} = 1/R_f + 1/R_s$, $\nu_2$ follows from
\begin{eqnarray}
\nu_2 B_z \Delta T = V_A + V_ B,  \label{nu2_1}
\end{eqnarray}
where  
\begin{eqnarray}
V_A &=&    I_{circ}^{x} R_{tot} \theta_{\rm H, s},  \label{nu2_2} \\
V_B &=& - I_{circ}^{x} R_{tot} \theta_{\rm H, f}.    \label{nu2_3}
\end{eqnarray}
Thus, the Nernst voltages are derived from the circulating current of $I_{circ}^{x}$; the total voltage is the series circuit. 
It should be noted that for the additional voltage (Eq.\ (\ref{nu2_1})) to appear, both a difference in the Seebeck coefficient ($S_f -S_s$) and a difference in the Hall angle ($\theta_{\rm H, s} -\theta_{\rm H, f}$) of the film and the substrate are necessary.

Fig.\ \ref{Fig_sche}(c) shows a scheme of the circuit, corresponding to Eqs. (\ref{nu2_1})-(\ref{nu2_3}).
Since $\nu_2$ is the contribution from the series circuit of Nernst voltages derived from the circulating current, which is essentially different from $\nu_1$, we call this contribution {\it the cooperative effect}.
Here, the sign difference with Eqs.\ (\ref{nu2_2}) and (\ref{nu2_3}) is due to the difference that the direction of the circulating current is positive in the $x$ direction for the ﬁlm but negative for the substrate, as shown in Fig.\ \ref{Fig_sche}(a).           
Since the amplitude and the sign of these contributions ($\nu_1$ and $\nu_2$) depend on the systems, we discuss specific examples in the next section.

In this section, only the Nernst coefficient was considered, however, we unambiguously obtained that this cooperate effect is present in the Hall coefficient ($R_{\rm H, \rm tot}$), the Seebeck coefficient with the anomalous components ($S_{\rm tot}^{AN}$), and the anomalous Nernst coefficient ($N_{\rm tot}$).

Finally we shortly comment on the anomalous Nernst effect in the bilayer system of ref.\ \cite{2024Zhu}.
When one neglects the Hall angle of the substrate, the second term of Eq.\ (\ref{N_tot}) is proportional to $(S_f -S_s) \rho^{xy}_{f}$, which is proportional to the circulating current.
Thus, this term includes the cooperative effect.   
This effect is the central part in the field of the anomalous Nernst effect based on the closed circuit, consisting of a magnetic material and a thermoelectric material \cite{2021Zhou,2021Yamamoto,2022Zhou,2023Zhou}.

\begin{figure}[h]
\begin{center}
\rotatebox{0}{\includegraphics[angle=0,width=1\linewidth]{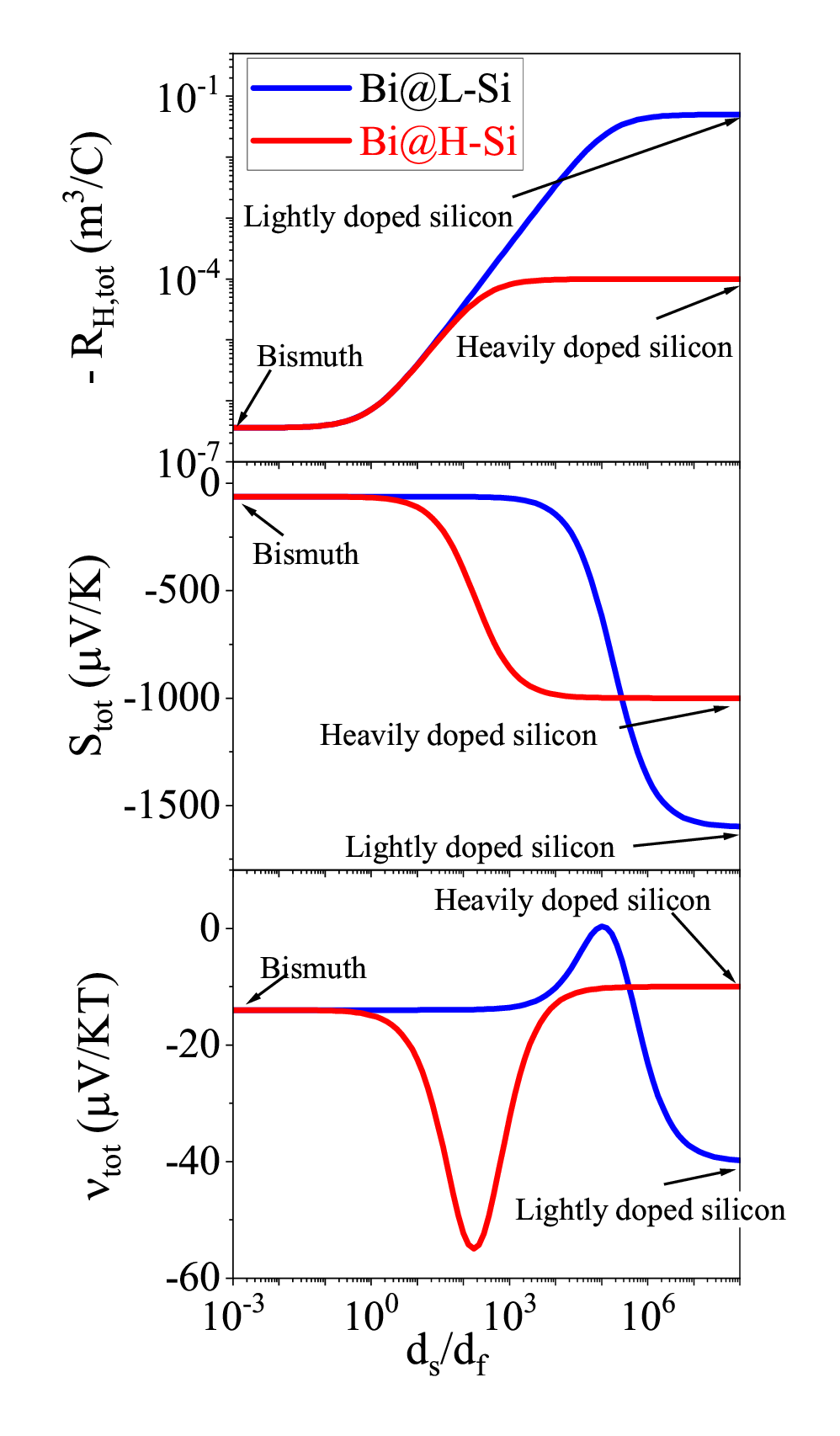}}
\caption{ Ratio of the thickness ($d_s/d_f$) dependence on the Hall coefficient, Seebeck coefficient and Nernst coefficient in bismuth film-Lightly doped silicon substrate system (Bi@L-Si) and bismuth film-heavily doped silicon substrate (Bi@H-Si). $d_f$ and $d_s$ indicates the thickness of film (bismuth) and substrate (silicon). }
\label{fig_exp}
\end{center}
\end{figure}
\begin{figure}[h]
\begin{center}
\rotatebox{0}{\includegraphics[angle=0,width=1\linewidth]{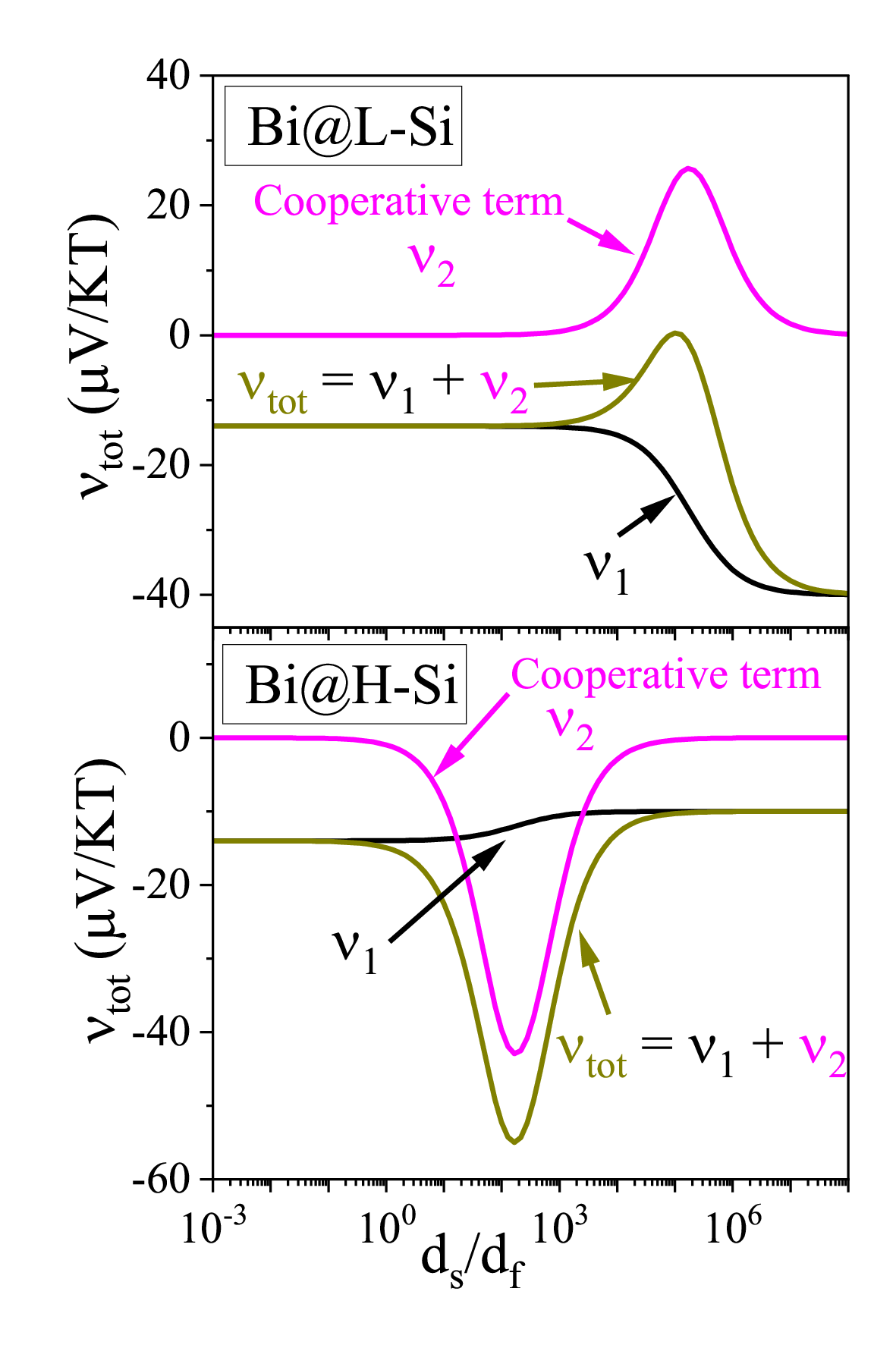}}
\caption{ $\epsilon_{\sigma}$ dependence of $\nu_1$ (black line), and $\nu_2$ (pink line) in bismuth film-lightly doped silicon substrate system (Bi@L-Si) and bismuth film-heavily doped silicon substrate system (Bi@H-Si), respectively. The blue and red lines indicate the total of Nernst coefficient ($\nu_{\rm tot}$).}
\label{fig_exp_Si}
\end{center}
\end{figure}

\section{Prediction of cooperative effect} 
To verify realistic amplitudes of this effect, the case of bismuth films on doped silicon substrates is discussed.
Here, we estimate the Hall coefficient, Seebeck coefficient, and Nernst coefficient from the experimental results.
Table \ref{Table1} shows experimental data at room temperature, extracted from refs. \cite{1954Morin,1955Geballe,1974Dunstan,2018Arisaka} for doped silicon and bismuth.
\begin{widetext}
\begin{table*}[t] 
\caption{Typical values of electrical conductivity, Hall coefficient, Hall mobility, Seebeck coefficient and Nernst coefficient at room temperatures. 
The experimental data are extracted from refs. \cite{1954Morin,1955Geballe,1974Dunstan,2018Arisaka}. \label{Table1}}
\begin{tabular}{c|c|c|c}
\hline
& lightly doped silicon & heavily doped silicon & bismuth \\  \hline \hline
Impurity concentration (cm$^{-3}$) &  3 $\times$10$^{14}$ & 1 $\times$ 10$^{17}$ &   \\ 
& (Sample num. 131 \cite{1954Morin})   &( Sample num. 139 \cite{1954Morin})   &  \\ 
& (Sample num.  2   \cite{1974Dunstan})   &(Sample num.  13 \cite{1974Dunstan})  &  \\ 
Electrical conductivity ($\mu\Omega^{-1}$m$^{-1}$) & 5 $\times$ 10$^{-6}$  \cite{1954Morin} & 0.005 \cite{1954Morin} & 0.88 \cite{2018Arisaka} \\
Hall coefficient (m$^3$/C) & -0.05 \cite{1954Morin} & -10$^{-4}$ \cite{1954Morin} & -3.6 $\times$ 10$^{-7}$ \cite{2018Arisaka} \\
Seebeck coefficient ($\upmu$V/K) & -1600 \cite{1955Geballe} & -1000 \cite{1955Geballe} & -62 \cite{2018Arisaka}\\
Nernst coefficient ($\upmu$V/KT) & -40 \cite{1974Dunstan} & -10 \cite{1974Dunstan} & -14 \cite{2018Arisaka} \\
\hline
\end{tabular} 
 \end{table*}
\end{widetext}

We set $d_f$ as the thickness of bismuth film, and $d_s$ as the thickness of lightly (heavily) doped silicon. 
Figure \ref{fig_exp} depicts the $d_s/d_f$ dependence of the Hall coefficient, Seebeck coefficient and Nernst coefficient in the bismuth film-lightly doped silicon substrate system (blue line) and bismuth film-heavily doped silicon substrate system (red line), respectively.
From here, we refer the bismuth film-lightly doped silicon substrate system as Bi@L-Si and bismuth film-heavily doped silicon substrate system as Bi@H-Si, respectively.
Although the Hall coefficient has a potential to have a peak due to the cooperative effect, the Hall and Seebeck coefficients change gradually in both Bi@L-Si and Bi@H-Si, as the ratio of thickness increases. 
On the other hand,  as already demonstrated in Fig.\ \ref{Fig_MC1_Nernst}, the Nernst coefficient exhibits a large peak of -55 $\upmu$V/KT unexpectedly at $d_s/d_f \simeq 100$ in Bi@H-Si, and in the case of Bi@L-Si, the Nernst coefficient decreases drastically at $d_s/d_f \simeq 10^{5}$.

In order to understand the origin and sign of the peak in Bi@L-Si and Bi@H-Si, $\nu_1$ and $\nu_2$ (Eq.\ (\ref{nu_divide})),
Figure \ref{fig_exp_Si} shows the $\epsilon_{\sigma}$ dependence of $\nu_1$, $\nu_2$, and $\nu_{\rm tot}$ in Bi@L-Si and  Bi@H-Si.
The cooperative effect ($\nu_2$) exhibits a negative peak at $d_s/d_f \simeq 100$ in Bi@H-Si, and a positive one at $d_s/d_f \simeq 10^{5}$ in Bi@L-Si.  
The sign of the cooperative effect depends on the material parameters.

As discussed in the introduction, a huge anomalous Nernst effect was observed in the bilayer system of Fe-Ga and n-type silicon substrate \cite{2024Zhu}. 
Although we did not consider the anisotropy explicitly and neglected interfacial properties, we obtained significant peaks in the system of bismuth thin films on different silicon substrates, depending on the film thickness.
 
\section{Conclusion}
In conclusion, we studied the thickness dependence of longitudinal and transverse responses in film-substrate systems in  a unified and general manner based on the parallel circuit model in matrix form.
We found that in transverse responses such as the Hall coefficient, the Nernst effect and the anomalous Nernst effect, there is a potential to obtain huge responses by adjusting the thickness of the film or the substrate, thereby fine-tuning the conductance ratio $\epsilon_\sigma$.
Our work emphasizes the importance of the cooperative effect in film and substrate systems.
Finally, in order to verify the role of the cooperative effect in realistic scenarios, bismuth film-doped silicon substrate systems were considered.
Using experimental parameters for the transport properties of bismuth and silicon, the Nernst effect of the system is drastically enhanced, originating from the interplay of film and substrate via the {\it cooperative effect}.

\begin{acknowledgments}
H.\ M.\ is grateful to M.\ Ogata for the discussion.
This work is supported by Grants-in-Aid for Scientific Research from the Japan Society for the Promotion of Science (No.\ JP20K03802, No.\ JP22K18954, and No.\ JP22KK0228), and JST-Mirai Program Grant (No.\ JPMJMI19A1).
\end{acknowledgments}

\appendix

\section{On the definitions of Hall and Nernst angles}
In the absence of a temperature gradient, zero heat current, and also zero electrical current in the $y$ direction, the electric field arising in the $y$ direction, when an electric field is applied in the $x$ direction, is obtained from
\begin{eqnarray}
\begin{pmatrix}
j^x_{\rm e,A}  \\
0  \\
\end{pmatrix}
= 
\bar{L}_{11,{\rm A}} 
\begin{pmatrix}
E_{x,{\rm A}}  \\
E_{y,{\rm A}}  \\
\end{pmatrix}
.
\end{eqnarray}
The ratio of $E_y$ and $E_x$, namely {\rm Hall angle}, is 
\begin{eqnarray}
\tan{\theta_{H,{\rm A}}} \simeq  \theta_{H,{\rm A}} = \frac{E_y}{E_x} =  -\frac{L_{11,{\rm A}}^{yx}}{L_{11,{\rm A}}^{yy}} = \frac{L_{11,{\rm A}}^{xy}}{L_{11,{\rm A}}^{yy}},   
\end{eqnarray}
where we assumed that the  Hall angle is much smaller than $\pi/2$.
The Hall mobility is defined as
\begin{eqnarray}
\mu_{H, {\rm A}} =  \frac{L_{11,{\rm A}}^{xy}}{L_{11,{\rm A}}^{yy}B_{z,{\rm A}}} = \frac{\sigma_{{\rm A}}^{xy}}{\sigma_{{\rm A}}^{yy}B_{z,{\rm A}} }.   
\end{eqnarray}

The Peltier angle is also obtained by the ratio of $(\nabla T)_x$ and $(\nabla T)_y$ as
\begin{eqnarray}
\tan{\alpha_{N,{\rm A}}} \simeq  \alpha_{N,{\rm A}} = \frac{(\nabla T)_y}{(\nabla T)_x}  = \frac{L_{12,{\rm A}}^{xy}}{L_{12,{\rm A}}^{yy}}.
\end{eqnarray}
where we also assumed that the Peltier angle is much smaller than $\pi/2$. 

\section{Hall resistivity in film-substrate system}
The electrical resistivity in system A is defined as
\begin{eqnarray}
 \begin{pmatrix}
E_{x,A}  \\
E_{y,A}  \\
\end{pmatrix}
= 
 \begin{pmatrix}
\rho^{xx}_A &  \rho^{xy}_{A} \\
\rho^{yx}_{A} &  \rho^{yy}_{A} \\
\end{pmatrix}
 \begin{pmatrix}
j_x^{e,A}  \\
j_y^{e,A}  \\
\end{pmatrix}
.  
\end{eqnarray}
By comparison with eq.\ (\ref{hall_2}), the Hall resistivity $\rho^{xy}_A$ is given as
\begin{eqnarray}
\rho^{xy}_A = \frac{-L^{xy}_{11,A}}{L^{xx}_{11,A}L^{yy}_{11,A} + (L^{xy}_{11,A})^2 }.
\end{eqnarray}
The Hall resistivity in the film-substrate system becomes
\begin{eqnarray}
\rho^{xy}_{\rm tot} &=& \frac{ -L^{xy}_{11,\rm tot} }{L^{xx}_{11,\rm tot} L^{yy}_{11,\rm tot} + (L^{xy}_{11,\rm tot})^2 } \nonumber \\
&=& \frac{1}{\frac{d_f}{d_f + d_s} L_{11,f}^{xx} } \frac{ \theta_{H,\rm tot} }{ (1 + \epsilon_{\sigma})(1 + \theta_{H,\rm tot}^2 ) }. \label{eq_rho_xy}
\end{eqnarray}
Substituting eq.\ (\ref{theta_tot}) to eq.\ (\ref{eq_rho_xy}), and approximating with $1 + \theta_{H,\rm tot}^2 \simeq 1$, the Hall resistivity is given as
\begin{eqnarray}
\rho^{xy}_{\rm tot} \simeq \frac{1}{\frac{d_f}{d_f + d_s} L_{11,f}^{xx} }\frac{\theta_{\rm H, f } + \epsilon_\sigma \theta_{\rm H, s }}{(1 + \epsilon_\sigma)^2}.
\end{eqnarray}

\section{n-layer system}
As discussed in Eq.\ (\ref{eq_sf_1}), the total response coefficient in the n-layer systems is given as
\begin{eqnarray}
\bar{L}_{\rm tot}^{{\rm n-layer}} \equiv \sum_{i=1}^{n}\frac{d_i}{d_{\rm tot}}\bar{L}_{i}. \label{eq_N_1}
\end{eqnarray}
where $d_i$ indicates the thickness of the i-th layer, and $d_{\rm tot}$ is the total thickness defined as $d_{\rm tot} = \sum_{i=1}^{n} d_i$. 
Here we define the ratio corresponding to Eq.\ (\ref{Riss_define}) as 
\begin{eqnarray}
\epsilon_{i} = \frac{L_{11, i}^{xx}d_i}{L_{11, 1}^{xx}d_1} = \frac{R_1}{R_i}.
\end{eqnarray}
where $R_i$ and $L_{11, i}^{xx}$ are the electrical resistivity in the i-th layer, and the electrical conductivity in the $i$-th layer, respectively.
Applying Eq.\ (\ref{eq_N_1}) to Eqs.\ (\ref{Hall_3}), (\ref{Seebeck_1}), (\ref{theta_1}) and (\ref{alpha_1}), the Hall coefficient, the Seebeck coefficient, the Hall angle and the Peltier angle become  
\begin{eqnarray}
R_{H, \rm tot}^{{\rm n-layer}} &=& \frac{\sum_{i=1}^n \frac{d_{\rm tot}}{d_i} \epsilon_{i}^2 R_{H,i} }{(\sum_{i}\epsilon_{i})^2 }, \\
S_{\rm tot}^{{\rm n-layer}} &=& \frac{\sum_{i=1}^n \epsilon_{i} S_i }{\sum_{i=1}^n \epsilon_{i} }, \\
\theta_{H,{\rm tot}}^{{\rm n-layer}} &=& \frac{\sum_{i=1}^n \epsilon_{i} \Theta_{H,i} }{\sum_{i=1}^n \epsilon_{i} }, \\
\alpha_{N,{\rm tot}}^{{\rm n-layer}} &=& \frac{\sum_{i=1}^n \epsilon_{i} S_{i}\alpha_{n,i} }{\sum_{i=1}^n S_i \epsilon_{i} },
\end{eqnarray}
Using these equations, the Nernst coefficient, the anomalous Nernst coefficient, and the Seebeck coefficient with the anomalous components are given as 
\begin{widetext}
\begin{eqnarray}
\nu_{\rm tot}^{{\rm n-layer}} &=& \frac{\sum_{i=1}^{n}\epsilon_{i}(\nu_{i} - S_i\theta_{H,i}/B_z)}{\sum_{i=1}^n \epsilon_{i} } + S_{\rm tot}^{{\rm n-layer}} \theta_{H, {\rm tot}}^{{\rm n-layer}}/B_z, \label{Nlayer_S}\\
S_{\rm tot}^{AN,{\rm n-layer}} &=& \frac{\sum_{i=1}^{n}\epsilon_{i}(S_{i}^{AN} - S_i\alpha_{{\rm n},i})}{\sum_{i=1}^n \epsilon_{i} } + S_{\rm tot}^{{\rm n-layer}} \alpha_{{\rm n}, {\rm tot}}^{{\rm n-layer}}, \label{Nlayer_S_an}\\
N_{\rm tot}^{{\rm n-layer}} &=&  \frac{\sum_{i=1}^{n}\epsilon_{i}(\nu_{i} - S_i\theta_{H,i})}{\sum_{i=1}^n \epsilon_{i} } + S_{\rm tot}^{{\rm n-layer}}\theta_{H, {\rm tot}}^{{\rm n-layer}}.\label{Nlayer_N_an}
\end{eqnarray}
\end{widetext}


\begin{thebibliography}{9}        


%
\bibitem{2020Nandihalli}
N.\ Nandihalli, C.\ J.\ Liu, T.\ Mori, Polymer based thermoelectric nanocomposite materials and devices: Fabrication and characteristics, Nano Energy {\bf 78}, 105186 (2020).

\bibitem{2019Mizuguchi}
Masaki Mizuguchi and Satoru Nakatsuji, Energy-harvesting materials based on the anomalous Nernst effect, Sci. Technol. Adv. Mater. {\bf 20}, 262 (2019).

\bibitem{2019Hinterleitner}
B.\ Hinterleitner, I.\ Knapp, M.\ Poneder, Yongpeng Shi, H.\ M\u{u}ller, G.\ Eguchi, C.\ Eisenmenger-Sittner, M.\ Stu{o}ger-Pollach, Y.\ Kakefuda, N.\ Kawamoto, Q.\ Guo, T.\ Baba, T.\ Mori, Sami Ullah, Xing-Qiu Chen and E.\  Bauer, Thermoelectric performance of a metastable thin-film Heusler alloy, Nature {\bf 576}, 85 (2019).

\bibitem{2022Garmroudi}
Fabian Garmroudi, Michael Parzer, Alexander Riss, Simon Beyer, Sergii Khmelevskyi, Takao Mori, Michele Reticcioli, Ernst Bauer, Large thermoelectric power factors by opening the band gap in semimetallic Heusler alloys, Materials Today Physics {\bf 27} 100742 (2022).

\bibitem{2023Bourgault}
D.\ Bourgault, H.\ Hajoum, R.\ Haettel, and E.\ Alleno, Unlocking microwatt power: enhanced performance of Fe–V–Al thin films in thermoelectric microgenerators,  Journal of Materials Chemistry A, {\bf 11} 19556 (2023)

\bibitem{2023Fujimoto}
Takuya Fujimoto, Masashi Mikami, Hidetoshi Miyazaki, Yoichi Nishino, Journal of Alloys and Compounds {\bf 969}, 172345 (2023).


\bibitem{2018Skinner}
B.\ Skinner and L.\ Fu, Large, nonsaturating thermopower in a quantizing magnetic field, Sci.\ adv.\ {\bf 4}, eaat2621 (2018). 

\bibitem{2022Mizoguchi}
Tomonari Mizoguchi, Hiroyasu Matsuura, Masao Ogata, Thermoelectric transport of type-I, II, and III massless Dirac fermions in a two-dimensional lattice model  Physical Review B {\bf 105}, 205203 (2022).

\bibitem{2022Hosoi}
Masashi Hosoi, Ikuma Tateishi, Hiroyasu Matsuura, Masao Ogata,Thin films of topological nodal line semimetals as a candidate for efficient thermoelectric converters, Physical Review B {\bf 105}, 085406 (2022).


\bibitem{2023Garmroudi}
Fabian Garmroudi, Michael Parzer, Alexander Riss, Cédric Bourgès, Sergii Khmelevskyi, Takao Mori, Ernst Bauer, Andrej Pustogow, High thermoelectric performance in metallic NiAu alloys via interband scattering.Sci.\ Adv.\ {\bf 9},\ eadj1611 (2023).

\bibitem{2015Ang}
Ran Ang, Atta Ullah Khan, Naohito Tsujii, Ken Takai, Ryuhei Nakamura, and Takao Mori, Thermoelectricity Generation and Electron–Magnon Scattering in a Natural Chalcopyrite Mineral from a Deep-Sea Hydrothermal Vent, Angewandte Chemie International Edition {\bf 54}, 12909 (2015).

\bibitem{2021Matsuura}
Hiroyasu Matsuura, Masao Ogata, Takao Mori, and Ernst Bauer, Theory of huge thermoelectric effect based on a magnon drag mechanism: Application to thin-film Heusler alloy, Phys.\ Rev.\ B {\bf 104}, 214421 (2021)

\bibitem{2019Tsujii}
N.\ Tsujii, A.\ Nishide, J.\ Hayakawa, T.\ Mori, Observation of enhanced thermopower due to spin fluctuation in weak itinerant ferromagnet,  Science Advances {\bf 5}, (2019).

\bibitem{2022Endo}
J.\ Endo, H.\ Matsuura, M.\ Ogata, Effect of paramagnon drag on thermoelectric transport properties: Linear response theory,  Phys.\ Rev.\ B {\bf 105}, 045101 (2022).


\bibitem{2003Delatorre}
R.\ Delatorre, M.\ Sartorelli, A.\ Schervenski, A.\ Pasa, and S.\ G\"{u}ths, Thermoelectric properties of electrodeposited CuNi alloys on Si, J. Appl. Phys. {\bf 93}, 6154 (2003).

\bibitem{2017Yordanov}
P.\ Yordanov, P.\ Wochner, S.\ Ibrahimkutty, C.\ Dietl, F.\ Wrobel, R.\ Felici, G.\ Gregori, J.\ Maier, B.\ Keimer, and H.\-U.\ Habermeier, Perovskite substrates boost the thermopower of cobaltate thin films at high temperatures, Appl.\ Phys.\ Lett.\ {\bf 110}, 253101 (2017).

\bibitem{2019Shimizu}
S.\ Shimizu, J.\ Shiogai, N.\ Takemori, S.\ Sakai, H.\ Ikeda, R.\ Arita, T.\ Nojima, A.\ Tsukazaki, and Y.\ Iwasa, Giant thermoelectric power factor in ultrathin FeSe superconductor, Nat.\ Commun.\ {\bf 10}, 1 (2019).

\bibitem{2019Zhang}
J.\ Zhang, W.\ Song, X.\ Ge, Z.\ Luo, and S.\ Yue, Boosted thermoelectric properties of molybdenum oxide thin films deposited on Si substrates, Mod.\ Phys.\ Lett.\ B {\bf 33}, 1950016 (2019).

\bibitem{2019Byeon}
D.\ Byeon, R.\ Sobota, K.\ Delime-Codrin, S.\ Choi, K.\ Hirata, M.\ Adachi, M.\ Kiyama, T.\ Matsuura, Y.\ Yamamoto, and M.\ Matsunami et al., Discovery of colossal Seebeck effect inmetallic Cu$_2$Se, Nat.\ Commun.\ {\bf 10}, 1 (2019).

\bibitem{2023matsubara}
Manaho Matsubara, Takahiro Yamamoto, and Hidetoshi Fukuyama, Two-band Model with High Thermoelectric Power Factor and Its Application to FeSe Thin Film, J.\ Phys.\ Soc.\ Jpn.\ {\bf 92}, 104704 (2023).





\bibitem{2000Young}
D. L. Young; T. J. Coutts; V. I. Kaydanov, Density-of-states effective mass and scattering parameter measurements by transport phenomena in thin films, Rev. Sci. Instrum. {\bf 71}, 462–466 (2000) .  
%
\bibitem{2002Young}
David L.\ Young, Helio Moutinho, Yanfa Yan, Timothy J.\ Coutts, Growth and characterization of radio frequency magnetron sputter-deposited zinc stannate, thin films, J. Appl. Phys. {\bf 92}, 310–319 (2002) .  
%
\bibitem{2003Young}
D. L. Young; J. F. Geisz; T. J. Coutts, Nitrogen-induced decrease of the electron effective mass in GaAs$_{1-x}$N$_x$ thin films measured by thermomagnetic transport phenomena, Appl. Phys. Lett. {\bf 82}, 1236–1238 (2003). 
%


%

\bibitem{2017Chuang}
T.\ C.\ Chuang, P.\ L.\ Su, P.\ H.\ Wu, and S. Y. Huang, Enhancement of the anomalous Nernst effect in ferromagnetic thin films, Rev.\ Phys.\ Rev.\ B {\bf 96}, 174406 (2017) .  
%
\bibitem{2018Hu}
Junfeng Hu, Benedikt Ernst, Sa Tu, Marko Kuve\v{z}di\'{c}, Amir Hamzi\'{c}, Emil Tafra,
Mario Basleti\'{c}, Youguang Zhang, Anastasios Markou, Claudia Felser, Albert Fert,
Weisheng Zhao, Jean-Philippe Ansermet, and Haiming Yu, Anomalous Hall and Nernst Effects in Co$_2$TiSn and Co$_2$Ti$_{0.6}$V$_{0.4}$Sn Heusler Thin Films, Rev.\ Phys.\ Rev.\ Applied {\bf 10}, 044037 (2018) .  
%

\bibitem{2020Park}
Gyu-Hyeon Park, Helena Reichlova, Richard Schlitz, Michaela Lammel, Anastasios Markou, Peter Swekis,
Philipp Ritzinger, Dominik Kriegner, Jonathan Noky, Jacob Gayles, Yan Sun, Claudia Felser, Kornelius Nielsch,
Sebastian T. B. Goennenwein, and Andy Thomas, Thickness dependence of the anomalous Nernst effect and the Mott relation of Weyl semimetal Co$_2$MnGa thin films, Rev.\ Phys.\ Rev.\ B {\bf 101}, 060406(R) (2020) .  
%


\bibitem{2022Yamazaki}
Takumi Yamazaki, Takeshi Seki, Rajkumar Modak, Keita Nakagawara, Takamasa Hirai ,
Keita Ito, Ken-ichi Uchida, and Koki Takanashi, Thickness dependence of anomalous Hall and Nernst effects in Ni-Fe thin films, Rev.\ Phys.\ Rev.\ B {\bf 105}, 214416 (2022) .  
%

\bibitem{2021Zhou}
W.\ Zhou, K.\ Yamamoto, A.\ Miura, R.\ Iguchi, Y.\ Miura, K.\ Uchida, Y.\ Sakuraba, Seebeck-driven transverse thermoelectric generation, Nat.\ Mater.\  {\bf 20}, 463 (2021).  
\bibitem{2021Yamamoto}
K.\ Yamamoto, R.\ Iguchi, A.\ Miura, W.\ Zhou, Y.\ Sakuraba, Y.\ Miura, K.\ Uchida, Phenomenological analysis of transverse thermoelectric generation and cooling performance in magnetic/thermoelectric hybrid systems, J.\ Appl.\ Phys.\ {\bf 129}, 223908 (2021).  
\bibitem{2022Zhou}
W.\ Zhou, T.\ Hirai, K.\ Uchida, Y.\ Sakuraba, Seebeck-driven transverse thermoelectric generation in on-chip devices, J.\ Phys.\ D: Appl.\ Phys.\ {\bf 55}, 335002 (2022).
\bibitem{2023Zhou}
W. Zhou, A. Miura, T. Hirai, Y. Sakuraba, K. Uchida, Seebeck-driven transverse thermoelectric generation in magnetic hybrid bulk materials, Appl.\ Phys.\ Lett.\ {\bf 122}, 062402 (2023)  


\bibitem{2015Quintana}
J.\ Alvarez-Quintana, Impact of the substrate on the efficiency of thin film thermoelectric technology, Appl.\ Therm.\ Eng.\ {\bf 84}, 206 (2015).

\bibitem{2023Riss}
A.\ Riss, M.\ St\"{o}ger, M.\ Parzer, F.\ Garmroudi, N.\ Reumann, B.\ Hinterleitner, T.\ Mori,
and E.\ Bauer, Criteria for Erroneous Substrate Contribution to the Thermoelectric
Performance of Thin Films, Phys. Rev. Applied {\bf 19}, 054024 (2023).  


\bibitem{1991Bergman}
D.\ J.\ Bergman and O.\ Levy, Thermoelectric properties of a composite medium, J.\ Appl.\ Phys.\ {\bf 70}, 6821 (1991).
\bibitem{1999Bergman}
D.\ J.\ Bergman and L.\ G.\ Fel, Enhancement of thermoelectric power factor in composite thermoelectrics, J.\ Appl.\ Phys.\ {\bf 85}, 8205 (1999).


\bibitem{2024Riss}
A.\ Riss,  F.\ Garmroudi, M.\ Parzer, A.\ Pustogow, T.\ Mori, and E.\ Bauer, Thermoelectric power factor of composites, Phys. Rev. Applied {\bf 21}, 014002 (2024).  


\bibitem{2024Zhu}
Weinan Zhou, Taisuke Sasaki, Ken-ichi Uchida, and Yuya Sakuraba, Direct-Contact Seebeck-Driven Transverse
Magneto-Thermoelectric Generation in Magnetic/Thermoelectric Bilayers, Adv. Sci. 2308543 (2024).  

\bibitem{Ziman} J.\ M.\ Ziman, {\it{Electrons and Phonons}} (Oxford, 1960).  See Table12.2.





\bibitem{1954Morin}
F.\ J.\ Morin and J.\ P.\ Maita, Electrical Properties of Silicon Containing Arsenic and Boron, \ Phys.\ Rev.\  {\bf 96}, 28 (1954).  

\bibitem{1955Geballe}
T.\ H.\ Geballe and G.\ W.\ Hull, Seebeck Effect in Silicon, \ Phys.\ Rev.\  {\bf 98}, 940 (1955).  

\bibitem{1974Dunstan}
W.\ Dunstan and D.\ W.\ Sear, The Nernst-Ettingshausen coefficient in n-type
silicon,\ J.\ Phys.\ C: Solid State Phys.\ {\bf 7}, 157 (1974).  


\bibitem{2018Arisaka}
Taichi Arisaka, Mioko Otsuka, and Yasuhiro Hasegawa, Investigation of carrier scattering process in polycrystalline bulk bismuth at 300 K, J.\ Appl.\ Phys.\ {\bf 123}, 235107 (2018).  
\end{thebibliography}
\end{document}